\documentclass{article}
\usepackage{spconf,amsmath,graphicx}

\usepackage{hyperref}
\hypersetup{
    colorlinks=true,
    citecolor=blue,
}
\usepackage{colortbl}

\usepackage{cite}
\usepackage{amsmath}
\usepackage{url}
\usepackage{amsfonts}
\usepackage{multirow}
\usepackage{makecell}
\usepackage{booktabs} 
\usepackage{subcaption}

\usepackage{pifont}

\title{The VoiceMOS Challenge 2024: Beyond Speech Quality Prediction}
\name{
    \begin{tabular}{c}
    Wen-Chin Huang$^{1}$,
    Szu-Wei Fu$^{2}$,
    Erica Cooper$^{3}$,
    Ryandhimas E. Zezario$^{4}$, \\ 
    Tomoki Toda$^{1}$,
    Hsin-Min Wang$^{4}$,
    Junichi Yamagishi$^{5}$,
    Yu Tsao$^{4}$
    \end{tabular}
    }
\address{
    $^{1}$Nagoya University, Japan $^{2}$NVIDIA, Taiwan\\
    $^{3}$National Institute Of Information And Communications Technology, Japan\\
    $^{4}$Academia Sinica, Taiwan $^{5}$National Institute of Informatics, Japan
}

\begin{document}
\ninept
\maketitle
\begin{abstract}
We present the third edition of the VoiceMOS Challenge, a scientific initiative designed to advance research into automatic prediction of human speech ratings. There were three tracks. The first track was on predicting the quality of ``zoomed-in'' high-quality samples from speech synthesis systems. The second track was to predict ratings of samples from singing voice synthesis and voice conversion with a large variety of systems, listeners, and languages. The third track was semi-supervised quality prediction for noisy, clean, and enhanced speech, where a very small amount of labeled training data was provided. Among the eight teams from both academia and industry, we found that many were able to outperform the baseline systems. Successful techniques included retrieval-based methods and the use of non-self-supervised representations like spectrograms and pitch histograms. These results showed that the challenge has advanced the field of subjective speech rating prediction.
\end{abstract}
\section{Introduction}
\label{sec:intro}

We present the third edition of the VoiceMOS Challenge (VMC) series. Founded in 2022, the VMC aims to use standardized datasets in diverse and challenging domains to understand and compare prediction techniques for human ratings of speech, specifically those collected through a mean opinion score (MOS) test. The main motivation is to foster the development in automatic, reference-free, and data-driven speech quality assessment methods to overcome costly and time-consuming human listening tests, which are generally regarded as the gold standard for evaluating speech synthesized or processed through various technologies such as text-to-speech synthesis (TTS), voice conversion (VC), singing voice synthesis (SVS), singing voice conversion (SVC), and speech enhancement (SE) \cite{speech-evaluation-review}.

The first edition of the VMC \cite{voicemos2022} was presented as a special session at Interspeech 2022 and attracted 22 participating teams from academia and industry. The challenge had a main track for MOS prediction based on BVCC \cite{bvcc}, a large-scale dataset of 187 different English TTS and VC systems, and a secondary out-of-domain (OOD) track with a very small amount of Mandarin samples and ratings from a separate listening test. The challenge results revealed that while the top systems of both tracks obtained very high correlations (0.939 and 0.979, respectively), a closer analysis showed that unseen \textbf{speakers} and \textbf{systems} were more difficult to predict. Also, participants were interested in more varied types of speech audio.

The second edition of VMC \cite{voicemos2023} was presented as a special session at ASRU 2023, with 10 teams participating in the challenge. With the goal of general, zero-shot, out-of-domain MOS prediction, it collaborated with the Blizzard Challenge (BC) 2023 \cite{bc2023}, which focused on French TTS, and the Singing Voice Conversion Challenge (SVCC) 2023 \cite{svcc2023}. It also provided a third noisy and enhanced speech track. Unlike the first edition, no MOS-labeled audio data from the target domains were provided. The results showed that there was still a gap between supervised and zero-shot settings, and that most teams could not achieve consistent performance across all tracks using the same model and training data, indicating that general-purpose MOS prediction remained an open research problem.

In this year's VMC, we organized three tracks.
The first track aimed to predict the MOS ratings of a “zoomed-in” subset comprising the top systems in the BVCC dataset collected through a separate listening test \cite{range-equalizing-bias}. This track reflects the need of present-day speech synthesis researchers to compare only high-quality synthesis systems.
The second track was based on a new dataset containing samples and their ratings from singing voice synthesis and conversion systems. This track can be regarded as an extension of track 2 in VMC 2023, with a larger variety of systems, datasets, and languages.
The third track was on reducing the need for labeled data for training MOS prediction models for noisy, clean, and enhanced speech. We focused on a semi-supervised setting, where only a very small amount of labeled data was provided. The motivation was to reflect the recent trend of unsupervised speech quality estimation \cite{speechlmscore, ssl-mos-zero-shot, vqscore}. The description of each track will be presented in Section~\ref{sec:description}.

This year, in addition to the baseline systems we prepared for each track, we received eight submissions from both academic and industrial research institutions, as will be shown in Section~\ref{sec:participants}. From the results shown in Section~\ref{sec:results}, we are pleased to confirm that there was at least one team that outperformed the baseline in each track, thus the challenge indeed advanced the technology in this field. In addition, from the system description questionnaires, we identified several successful techniques used by top systems, including retrieval-based prediction, ensemble learning, and the use of features other than the mainstream self-supervised learning (SSL) representations, such as mel spectrograms and pitch histograms. These findings will be discussed in Section~\ref{sec:system-analysis}. Finally, the participants' fruitful feedback summarized in Section~\ref{sec:feedback} sheds light on the most urgent future directions pointed directly by the research community.

\section{Challenge Description}
\label{sec:description}

The challenge took place from April 10 to June 20, 2024.  The test audio samples for all tracks were released to participants on May 27.  Predictions were due on June 3, and the results were released to the participants on June 13. Each team was asked to submit a system description form.

\subsection{Tracks and datasets}

This year's challenge tracks included MOS prediction for ``zoomed-in'' systems, MOS prediction for singing voices, and semi-supervised MOS prediction for noisy, clean, and enhanced speech. Table \ref{tab:datasets} summarizes the datasets used in each track. The details are  described next. 

\begin{table*}[t]
	\centering
	\caption{Summary of the datasets for each track.}
	
	\centering
	\begin{tabular}{ c c c c c c c c c}
		\toprule
		\multirow{2}{*}[-1pt]{Track} & \multirow{2}{*}[-1pt]{Type} & \multicolumn{3}{c}{\# Samples} & \multicolumn{3}{c}{\# Systems (conditions)} & \multirow{2}{*}[-2pt]{\makecell{\# ratings\\per sample}}\\
		\cmidrule(lr){3-5} \cmidrule(lr){6-8}
		& & Train & Dev & Test & Train & Dev & Test & \\
		\midrule
		1 & Zoomed-in TTS & -- & 1000 & 1000 & -- & 88 & 25\%: 46; 12\%: 23 & Dev: 4; Test (25\%): 8; Test (12\%): 12 \\
		\midrule
		2 & Singing Voices & 2000 & 544 & 645 & 35 & 35 & 35 & 5\\
            \midrule
            3 & Noisy \& Enhanced & 60 & 40 & 280 & 5 & 5 & 7 & Train \& Dev: 8; Test:5\\
		\bottomrule
	\end{tabular}
	\label{tab:datasets}
\end{table*}

\subsubsection{Track 1: MOS prediction for ``zoomed-in'' systems}

This track is motivated by a real-world scenario where the researchers wish to evaluate a speech generation model under development, whose quality is expected to be better than any previous system. In such a case, the MOS predictor is required to correctly predict the subjective ranking between top-quality systems, with the possibility of not having seen samples from new systems before. To support this scenario, we utilized a new set of ``zoomed-in'' subjective ratings curated in a recent work \cite{range-equalizing-bias}. The authors ranked the 187 systems in the BVCC dataset based on their MOS ratings and created several subsets by selecting approximately 50\%, 25\%, 12\%, and 6\% of the highest-rated systems. Then, new separate listening tests were conducted for each new subset of systems. Note that the samples are identical to the original BVCC dataset (in 16kHz), and only the ratings are newly collected in that work.

In the training phase, to reflect the scenario described above, we did not provide any new training data and instructed participants to make use of the ratings in the original BVCC dataset. The test set consists of 1000 samples from the 25\% subset, 500 of which are also included in the 12\% subset. Then, for the validation subset, we selected 1000 samples from the 50\% subset that were present in neither the 25\% subset nor the 12\% subset. This design allowed us to investigate the performance of the participating systems w.r.t. different zoom-in rates.

Since the newly collected ratings were already publicized at the time we organized the challenge, we needed to take action to prevent participants from cheating. First, throughout the challenge, we did not inform the participants that the ratings were based on those collected in \cite{range-equalizing-bias}. Second, the number of samples in the validation and test sets, which was 1000, was chosen intentionally to not match any total size of the subsets in \cite{range-equalizing-bias} (which were 2976, 1472, and 736, respectively).

\subsubsection{Track 2: MOS prediction for singing voice}

Track 2 in VMC 2023 was based on the submissions of SVCC 2023. The dataset was based on the NUS-HLT Speak-Sing (NHSS) dataset \cite{nhss}, which only contained English singing voice samples from non-professional singers. These limitations of the SVCC 2023 dataset lead to the interest of developing a dataset for singing voice evaluation with a larger variety of systems, datasets, and languages.

In this track, we utilized a newly collected dataset named SingMOS \cite{singmos}. It is a compilation of natural singing voice samples, vocoder analysis-synthesis samples, and generated samples from SVS and SVC systems. The samples are in Chinese and Japanese, with a sampling rate of 16kHz. The datasets used in the training set include M4Singer, OpenCPop, Kiritan, and JVS-Music, with a total of 2000 samples. The validation set contains 500 samples from datasets and systems used in the training set, as well as 44 additional samples from two datasets not seen during training: Ofuton and Kising. The evaluation set contains 500 and 44 samples from datasets and systems in the training and validation sets, respectively, and 101 additional samples from unseen systems in the training and validation sets. Each sample was rated by five listeners.

\subsubsection{Track 3: Semi-supervised MOS prediction for noisy, clean, and enhanced speech}


Ensuring access to sizable training samples is crucial for deploying non-intrusive speech quality prediction models, but extensive training data is not always available. It is interesting to observe the potential robustness of the overall model performance when training the model with a \textbf{very limited} amount of data and testing it on the \textbf{entirely unseen} testing scenario. This consideration led us to deploy this track in the challenge.  

Instead of just measuring naturalness, we follow the subjective evaluation methodology in ITU-T P.835 \cite{p835} to assess speech signal quality (SIG), background intrusiveness (BAK), and overall quality (OVRL). The SIG metric assesses the level of speech distortion in an audio sample, asking listeners to evaluate the degree of distortion in the played speech sample. The highest score (5) indicates no distortion, while the lowest score (1) indicates severe distortion of the speech sample. Next, the BAK metric indicates the degree of background intrusiveness, which is related to how much background noise is contained in the speech sample. Similar to SIG, the scores range from 1 to 5, where the highest score (5) indicates that there is no background noise in the speech sample, and the lowest score (1) indicates that the speech sample is heavily contaminated by noise. In addition, the OVRL metric indicates the overall quality of the speech sample, with the highest score (5) indicating excellent quality and the lowest score (1) indicating poor quality. Similar to SIG and BAK, the scores range from 1 to 5. 

The training and validation sets were based on the samples submitted to the UDASE task of the 7th CHiME challenge \cite{chime7-udase, chime7-evaluation} (5 different conditions: including speech processed by 4 speech enhancement systems and unprocessed noisy speech). We randomly sampled 60 and 40 utterances with the corresponding SIG, BAK, and OVRL scores as the training and validation sets. Each sample was rated by 8 listeners.
The evaluation set was based on another dataset, the VoiceBank-DEMAND \cite{Voicebank_Demand} dataset. 
We specifically selected 40 clean utterances to construct the evaluation set. These 40 clean utterances were contaminated by one of four types of noise (car, white, music, and cafeteria) at an SNR of 0 dB to produce the corresponding noisy utterances, and each noisy utterance was processed by five speech enhancement systems (BSSE \cite{hung2022boosting}, MPSENet \cite{lu2023mp}, CMGAN \cite{cao2022cmgan}, DEMUCS \cite{defossez2020real}, and Wiener \cite{loizou2013speech}) to produce enhanced utterances. Finally, the evaluation set contains 280 utterances.
Ten listeners (six males and four females, all native English speakers) were invited to rate the speech samples. Each utterance was evaluated by five listeners.



\subsection{Challenge rules and phases}
\label{ssec:rules}

This year's challenge was held on CodaBench\footnote{\url{https://www.codabench.org/competitions/2650/}}, an advanced version of the previous CodaLab platform. It is an open-source web-based platform for machine learning competitions and reproducible research. The participants could choose to participate in any track as long as they submitted their answers in the evaluation phase.

To promote research reproducibility, we prohibited the use of in-house datasets unless the participants were willing to (1) publicize the datasets after the challenge, or (2) open-source their system, along with any ready-to-use model checkpoints. In addition, to enforce a complete semi-supervised setting for track 3, we further limited participants to use only the labeled data ($\langle$speech, subjective rating$\rangle$ pairs) provided by the organizers. This implies that other off-the-shelf subjective speech quality estimators (including but not limited to PESQ, STOI, and any other open-sourced data-driven quality estimators, such as SSL-MOS \cite{ssl-mos}) could not be used for system development and training.

In the training phase, we directly provided links for downloading the training and validation sets, including the speech samples and ratings. Track 1 was an exception, as we only provided instructions for obtaining the BVCC dataset. We also provided the set of sample IDs for the validation set. Participants were about to upload their validation results to the system and submit their scores to the leaderboard, such that they could get to know how their systems performed and see the results of other teams' submissions.  

For the evaluation phase, as all track 1 and 2 samples were accessible in the training phase, we released only the sample IDs in the evaluation set. For track 3, we provided both the samples and the IDs.
Participants were allowed to make up to ten submissions, and the leaderboard was only used for sanity checking on whether the sample IDs and samples matched. After completing the analysis of the submissions from participants, we returned to the participants (1) the results, (2) their unique team ID, and (3) the ground-truth labels to the evaluation set of each track.

\begin{table}[t]

    \centering
    \caption{List of participant affiliations, in random order.}
    \label{tab:teams}

    \begin{tabular}{ l c c c }
        \toprule
        \multirow{2}{*}[-1pt]{Affiliation} & \multicolumn{3}{c}{Track} \\
         & 1 & 2 & 3 \\
        \midrule
        Nankai University, China & Y & Y & Y \\
        University of Science and Technology of China, China & N & Y & N \\
        University of Tokyo, Japan & Y & Y & N \\
        Southern University of Science and Technology, China & Y & Y & Y \\
        University of West Bohemia, Czech Republic & Y & Y & Y \\
        Logitech & Y & Y & Y \\ 
        Constructor Technology & Y & N & N \\ 
        Codemill AB & Y & N & N \\ 
        \bottomrule
    \end{tabular}
\end{table}

\section{Participants and baseline systems}
\label{sec:participants}

We received evaluation phase submissions from five teams from academia and three teams from industry for eight in total from six different countries.
Table~\ref{tab:teams} briefly summarizes the participants, their affiliations, as well as the tracks they participated in. For each track, we have seven, six, and four participating teams, respectively.
Teams were randomly assigned anonymized numerical team IDs, and as organizers, we do not actively show the mapping from team IDs to affiliations. In the following sections, we will refer to each team with their ID.

We also have a baseline system for each track. For track 1, the SSL-MOS model \cite{ssl-mos}, which was also the baseline in VMC 2022 and 2023, was used. We directly used the source code and the provided model checkpoint as-is from the open-source repository\footnote{\url{https://github.com/nii-yamagishilab/mos-finetune-ssl/}}. The model was trained on the original BVCC dataset. For track 2, the SSL-MOS was also used, and it was trained on the provided training set.
For track 3, a recently proposed unsupervised speech quality predictor named VQScore was used \cite{vqscore}\footnote{\url{https://github.com/JasonSWFu/VQscore}}. The model was trained with the 460-hour LibriSpeech \cite{librispeech} train-clean set. The results of DNSMOS P.835 \cite{dnsmosp835} are also provided as a supervised baseline which is trained on 75 hours ⟨speech,
subjective rating⟩ data pairs.

\section{Results}
\label{sec:results}

\subsection{Evaluation metrics}

For tracks 1 and 2, following previous VMCs, we computed both utterance-level and system-level mean squared error (MSE), linear correlation coefficient (LCC), Spearman rank correlation coefficient (SRCC), and Kendall's Tau rank correlation coefficient (KTAU).  Our primary metric was system-level SRCC with KTAU as a secondary metric because, in a real-life MOS prediction task, we mainly want to know the {\em rankings} of the systems under consideration. For track 3, the utterance-level correlation is selected as the evaluation metric, following prior works on non-intrusive speech quality prediction models in denoising tasks \cite{dnmos, mosanet}. We calculated the utterance-level LCC of SIG, BAK, and OVRL. We considered a ``system'' as a unique combination of speech enhancement method, SNR, and noise type.

\subsection{Track 1 results}

\begin{table}[t]
\centering
\caption{Results for track 1 with different zoomed-in rates. Bold face indicates the best system w.r.t. each metric.}
\label{tab:track1}
\footnotesize
\begin{subtable}[t]{0.48\textwidth}
    \begin{tabular}{@{}crrrrrrrr@{}}
    \toprule
    \multicolumn{1}{l}{}     & \multicolumn{4}{c|}{Utterance-level}                                                                      & \multicolumn{4}{c}{System-level}                                                                                 \\ \cmidrule(l){2-9} 
    \multicolumn{1}{l}{}     & \multicolumn{1}{c}{MSE} & \multicolumn{1}{c}{LCC} & \multicolumn{1}{c}{SRCC} & \multicolumn{1}{c|}{KTAU}  & \multicolumn{1}{c}{MSE} & \multicolumn{1}{c}{LCC} & \multicolumn{1}{c}{SRCC} & \multicolumn{1}{c}{KTAU} \\ \midrule
    \multicolumn{1}{c|}{B01} & 1.154                   & 0.508                   & 0.509                    & \multicolumn{1}{r|}{0.358} & 0.998                   & 0.750                   & 0.745                    & 0.539                    \\ \midrule
    \multicolumn{1}{c|}{T01} & 0.358                   & 0.490                   & 0.491                    & \multicolumn{1}{r|}{0.342} & 0.162                   & 0.703                   & 0.700                    & 0.495                    \\
    \multicolumn{1}{c|}{T02} & 0.355                   & 0.550                   & 0.555                    & \multicolumn{1}{r|}{0.394} & 0.141                   & 0.789                   & 0.769                    & 0.578                    \\
    \multicolumn{1}{c|}{T03} & 0.406                   & 0.471                   & 0.488                    & \multicolumn{1}{r|}{0.344} & 0.200                   & 0.674                   & 0.672                    & 0.483                    \\
    \multicolumn{1}{c|}{T04} & 0.718                   & 0.421                   & 0.391                    & \multicolumn{1}{r|}{0.272} & 0.494                   & 0.728                   & 0.698                    & 0.512                    \\
    \multicolumn{1}{c|}{T05} & 0.287                   & \textbf{0.686}                   & \textbf{0.679}                    & \multicolumn{1}{r|}{\textbf{0.497}} & 0.073                   & \textbf{0.944}                   & 0.940                    & 0.773                    \\
    \multicolumn{1}{c|}{T06} & \textbf{0.255}                   & 0.658                   & 0.676                    & \multicolumn{1}{r|}{0.493} & \textbf{0.047}                   & 0.931                   & \textbf{0.943}                    & \textbf{0.793}                    \\
    \multicolumn{1}{c|}{T07} & 0.753                   & 0.377                   & 0.359                    & \multicolumn{1}{r|}{0.254} & 0.446                   & 0.724                   & 0.675                    & 0.493                    \\ \bottomrule
    \end{tabular}
    \caption{Zoomed-in rate: 25\%}
\end{subtable}

\begin{subtable}[t]{0.48\textwidth}
    \begin{tabular}{@{}crrrrrrrr@{}}
    \toprule
    \multicolumn{1}{l}{}     & \multicolumn{4}{c|}{Utterance-level}                                                                      & \multicolumn{4}{c}{System-level}                                                                                 \\ \cmidrule(l){2-9} 
    \multicolumn{1}{l}{}     & \multicolumn{1}{c}{MSE} & \multicolumn{1}{c}{LCC} & \multicolumn{1}{c}{SRCC} & \multicolumn{1}{c|}{KTAU}  & \multicolumn{1}{c}{MSE} & \multicolumn{1}{c}{LCC} & \multicolumn{1}{c}{SRCC} & \multicolumn{1}{c}{KTAU} \\ \midrule
    \multicolumn{1}{c|}{B01} & 0.741                   & 0.422                   & 0.417                    & \multicolumn{1}{r|}{0.285} & 0.589                   & 0.608                   & 0.609                    & 0.444                    \\ \midrule
    \multicolumn{1}{c|}{T01} & 0.296                   & 0.402                   & 0.421                    & \multicolumn{1}{r|}{0.286} & 0.127                   & 0.540                   & 0.513                    & 0.333                    \\
    \multicolumn{1}{c|}{T02} & 0.410                   & 0.520                   & 0.544                    & \multicolumn{1}{r|}{0.377} & 0.240                   & 0.745                   & 0.702                    & 0.467                    \\
    \multicolumn{1}{c|}{T03} & 0.264                   & 0.481                   & 0.514                    & \multicolumn{1}{r|}{0.357} & 0.082                   & 0.779                   & 0.768                    & 0.539                    \\
    \multicolumn{1}{c|}{T04} & 0.457                   & 0.311                   & 0.281                    & \multicolumn{1}{r|}{0.191} & 0.270                   & 0.571                   & 0.549                    & 0.364                    \\
    \multicolumn{1}{c|}{T05} & \textbf{0.203}                   & \textbf{0.654}                   & 0.658                    & \multicolumn{1}{r|}{0.464} & \textbf{0.056}                   & 0.834                   & 0.788                    & 0.610                    \\
    \multicolumn{1}{c|}{T06} & 0.221                   & 0.629                   & \textbf{0.695}                    & \multicolumn{1}{r|}{\textbf{0.505}} & 0.057                   & \textbf{0.890}                   & \textbf{0.952}                    & \textbf{0.824}                    \\
    \multicolumn{1}{c|}{T07} & 0.517                   & 0.300                   & 0.294                    & \multicolumn{1}{r|}{0.207} & 0.251                   & 0.676                   & 0.672                    & 0.444                    \\ \bottomrule
    \end{tabular}    
    \caption{Zoomed-in rate: 12\%}
\end{subtable}

\end{table}
\begin{figure}[t]
	\centering
	\includegraphics[width=\linewidth]{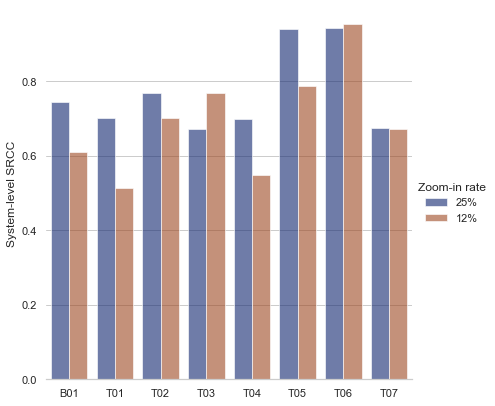}
	\caption{\label{fig:track1}Bar plot of system-level SRCC values of all participants in track 1.}	
\end{figure}

Table~\ref{tab:track1} shows the results from the baseline and all participating teams, and the bar plot of the main metric, system-level SRCC, is shown in Figure~\ref{fig:track1}. First, we found that given only the original BVCC labels, most systems (including the baseline) struggled to rank the systems in the 12\% zoom-in rate, except for T03 and T06. This result shows that the 12\% zoom-in rate labels are indeed harder to predict compared to 25\%. The baseline ranked 4th and 6th w.r.t. the 25\% and 12\% zoom-in rates, showing that participants have indeed advanced the prediction of zoomed-in systems. Remarkably, T05 and T06 were the top two systems in all eight metrics w.r.t. both 25\% and 12\% zoom-in rates. 


\begin{table}[t]
\centering
\caption{Most difficult 5 systems to predict for track 1 with different zoomed-in rates. The third column, \# teams, refers to the number of times it appeared in the top 5 most difficult system for each team.}
\label{tab:track1-difficult}

\centering
\begin{tabular}{@{}cccc@{}}
\toprule
\multicolumn{1}{l}{Zoom-in rate} & System name               & \# teams  & \makecell{Ranking among\\all systems} \\ \midrule
\multirow{5}{*}{25\%}            & BC2016-L       & 5        & 16/46                \\
                                 & BC2011-G       & 5        & 34/46                \\
                                 & BC2010-J       & 3        & 27/46                \\
                                 & VCC2020-T27    & 3        & 24/46                \\
                                 & BC2010-P       & 2        & 40/46                \\ \hline
\multirow{5}{*}{12\%}            & BC2010-A       & 6        & 7/23                 \\
                                 & BC2010-J       & 5        & 19/23                \\
                                 & BC2011-G       & 5        & 21/23                \\
                                 & BC2010-M       & 4        & 15/23                \\
                                 & VCC2016-source & 4        & 11/23                \\ \hline
\end{tabular}

\end{table}

\noindent\textbf{Difficult systems in track 1.} Here we analyzed which systems in track 1 were difficult to predict in the zoomed-in setting. With the predicted rankings produced by each participating team, we calculated the \textbf{``absolute ranking difference''} of each system. For instance, if BC2010-J ranks 19th out of a total of 23 systems, but a team predicts it as 9th, then BC2010-J has an absolute ranking difference of 10. Then, for each system, we counted how many times it appeared in the top 5 most difficult systems for each team. Table~\ref{tab:track1-difficult} shows the results.  Interestingly, we observed that none of the most difficult systems to rank were systems that were unseen with respect to the BVCC training data.  Also, we observed that at both of the zoom-in levels, three out of the five most difficult systems are hybrid TTS systems (using hidden Markov models or other data-driven machine learning based approaches to select units for concatenation).  Six out of the 23 total systems at the 12\% zoom level, and 11 out of 46 systems at the 25\% level, are known to be hybrid -- their concentration among the difficult systems indicates that this type of system may especially present a challenge for predictors.


\subsection{Track 2 results}

\begin{table}[t]
\centering
\caption{Results for track 2.}
\label{tab:track2}
\begin{tabular}{@{}crrrrrrrr@{}}
\toprule
\multicolumn{1}{l}{}     & \multicolumn{4}{c|}{Utterance-level}                                                                      & \multicolumn{4}{c}{System-level}                                                                                 \\ \cmidrule(l){2-9} 
\multicolumn{1}{l}{}     & \multicolumn{1}{c}{MSE} & \multicolumn{1}{c}{LCC} & \multicolumn{1}{c}{SRCC} & \multicolumn{1}{c|}{KTAU}  & \multicolumn{1}{c}{MSE} & \multicolumn{1}{c}{LCC} & \multicolumn{1}{c}{SRCC} & \multicolumn{1}{c}{KTAU} \\ \midrule
\multicolumn{1}{c|}{B01} & 0.419                   & 0.594                   & 0.605                    & \multicolumn{1}{r|}{0.442} & 0.079                   & 0.851                   & \textbf{0.859}                    & \textbf{0.687}                    \\ \midrule
\multicolumn{1}{c|}{T01} & 0.366                   & 0.605                   & 0.603                    & \multicolumn{1}{r|}{0.440} & \textbf{0.051}                   & 0.858                   & 0.837                    & 0.684                    \\
\multicolumn{1}{c|}{T03} & 0.432                   & 0.597                   & 0.583                    & \multicolumn{1}{r|}{0.426} & 0.061                   & 0.848                   & 0.819                    & 0.637                    \\
\multicolumn{1}{c|}{T04} & 0.363                   & 0.624                   & 0.604                    & \multicolumn{1}{r|}{0.445} & 0.056                   & \textbf{0.869}                   & 0.833                    & 0.640                    \\
\multicolumn{1}{c|}{T05} & 0.363                   & 0.609                   & 0.593                    & \multicolumn{1}{r|}{0.434} & 0.069                   & 0.791                   & 0.807                    & 0.657                    \\
\multicolumn{1}{c|}{T06} & \textbf{0.358}                   & \textbf{0.637}                   & \textbf{0.625}                    & \multicolumn{1}{r|}{\textbf{0.460}} & 0.063                   & 0.841                   & 0.831                    & 0.657                    \\
\multicolumn{1}{c|}{T08} & 0.384                   & 0.628                   & 0.620                    & \multicolumn{1}{r|}{0.455} & 0.072                   & 0.845                   & 0.856                    & 0.687                    \\ \bottomrule
\end{tabular}
\end{table}
\begin{figure}[t]
	\centering

        \includegraphics[width=0.8\linewidth]{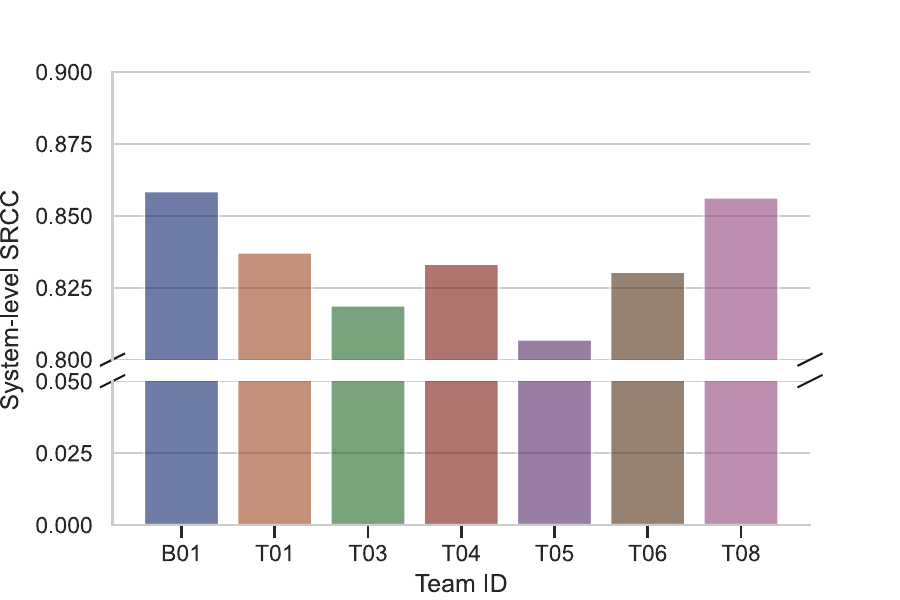}
 
	\caption{\label{fig:track2}Bar plot of system-level SRCC values of all participants in track 2.}	
\end{figure}

For track 2, Table~\ref{tab:track2} shows the results from the baseline and all participating teams, and the bar plot of the main metric, system-level SRCC, is shown in Figure~\ref{fig:track2}. If we only look at the system-level SRCC, unfortunately, no participant outperformed the baseline. Nonetheless, the baseline ranked first in only the system-level SRCC and system-level KTAU, and the T06 system ranked first in all four utterance-level metrics. Moreover, we observed that in Table~\ref{tab:track2}, the difference between the worst system and the top system in each metric was in fact small.
This indicates that the participants could not identify an effective method for this track.

\subsection{Track 3 results}

\begin{table}[t]
\centering
\caption{Results (LCC) for track 3.}
\label{tab:track3}
\begin{tabular}{@{}lrrr@{}}
\toprule
                         & \multicolumn{1}{c}{SIG } & \multicolumn{1}{c}{BAK } & \multicolumn{1}{c}{OVRL} \\ \midrule

\multicolumn{1}{c|}{DNSMOS P.835 (supervised) \cite{dnsmosp835}} & 0.225                       & 0.845                       & 0.638                        \\
\midrule
\multicolumn{1}{c|}{B01 (VQScore) (unsupervised) \cite{vqscore}} & 0.261                       & 0.737                       & 0.592                        \\
\midrule
\multicolumn{1}{c|}{T01} & 0.054                       & 0.758                       & 0.661                        \\
\multicolumn{1}{c|}{T03} & -0.116                      & 0.694                       & 0.254                        \\
\multicolumn{1}{c|}{T04} & 0.207                       & \textbf{0.867}                       & 0.711                        \\
\multicolumn{1}{c|}{T06} & \textbf{0.297}                       & 0.827                       & \textbf{0.713}                        \\ \bottomrule
\end{tabular}
\end{table}
\begin{figure}[t]
	\centering
	\includegraphics[width=0.8\linewidth]{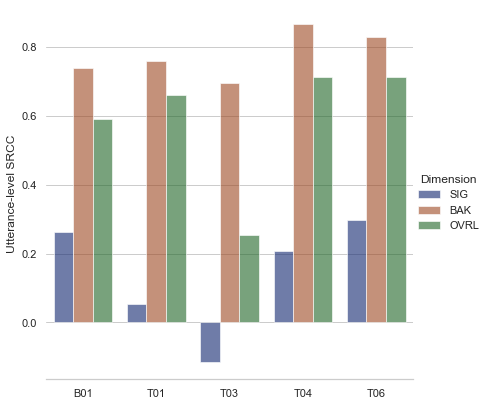}
	\caption{\label{fig:track3}Bar plot of utterance-level LCC values of SIG, BAK and OVRL of all participants in track 3.}	
\end{figure}

The overall evaluation results of track 3 are presented in Table~\ref{tab:track3}. Among the three metrics, SIG is the most difficult to predict, as indicated by lower correlation scores compared to BAK and OVRL. Based on the submitted systems, we noticed that all systems incorporated large pretrained models, either SSL models (T03, T04, T06) or Whisper (T01). Moreover, predicting SIG, BAK, and OVRL with very limited labeled training utterances is quite challenging, especially when dealing with unseen scenarios. In this context, directly fine-tuning SSL models with corresponding speech assessment labels using the regression loss between the estimated scores and the corresponding labels (T04, T06) showed better effectiveness than performing additional fine-tuning of SSL models with extra training data using the corresponding SSL loss (T03). We also observed that fusion estimation (combining estimated scores from two different modules) can provide better performance with limited training data, as indicated by the good performance of T06, where the predicted MOS is combined with the MOS retrieved from the datastore using k-nearest neighbors (kNN) method. Additionally, we noted that no single system excels across all metrics. Specifically, T06 performs best in predicting SIG and OVRL, while T04 performs best in predicting BAK. Interestingly, for BAK estimation, first fine-tuning the SSL model for SNR estimation and then fine-tuning it for BAK score estimation can significantly improve prediction performance. 

Comparing VQScore (B01) and the submitted systems to DNSMOS P.835, we observed that large-scale labeled training data may \textbf{not} be necessary to train a speech quality predictor. We also observed that the average standard deviation (std) among labelers in the evaluation set is 0.686, 0.571, and 0.559 for SIG, BAK, and OVRL, respectively. SIG has the highest std, implying it is difficult for labelers to obtain aligned ratings. However, although it was also found in \cite{chime7-evaluation} that SIG is the most difficult metric to predict, the underlying reason remains unclear.


\section{Analysis of the participating systems}
\label{sec:system-analysis}

We received system descriptions from participants except for T07, and we conducted an analysis of the submitted systems. Below we provide our findings and insights.

\subsection{Datasets}

For tracks 1 and 2 where we did not pose any constraint on the data usage, to our surprise, many teams did not use additional labeled datasets\footnote{Here, by saying ``additional labeled dataset'', we mean datasets with subjective ratings as labels. We do not take into account, for instance, data used for self-supervised pre-training.}. For track 1, four out of the six\footnote{T07 was not counted.} teams did not use additional labeled data, and for track 2, three out of six teams did not use additional labeled data. Popular additional dataset choices included SOMOS \cite{somos} and past Blizzard Challenges. T03 even tried to use the ConferencingSpeech 2022 training sets \cite{conferencingspeech2022}, which mainly consist of noisy speech samples and their ratings.

For track 3 where we did not allow using additional datasets with subjective labels, two out of the four participants did not use additional noise or speech data. Other participants used datasets like the AudioSet \cite{audioset}, DNS \cite{dns-is2020}, ESC-50 \cite{esc-50}, MS-SNSD \cite{ms-snsd}, and ASVSpoof 2019 \cite{asvspoof2019} to either pre-train the model or perform data augmentation.

\subsection{Features and modeling}

As in previous challenges, using self-supervised learning (SSL) representations still remained a popular approach, as almost all systems (including the baseline) used features from models including wav2vec 2.0 \cite{wav2vec2}, XLSR \cite{xlsr}, WavLM \cite{wavlm} and BYOL-S \cite{byol-s}. While previous challenge participants only used one or more SSL features, some participants this year attempted to use of other features. For instance, T01 used magnitude spectrograms,  mel spectrograms and Whisper features \cite{whisper}, T04 used features from HiFi-GAN discriminators \cite{hifigan}, T05 used mel spectrograms, and T08 used pitch histograms. On the other hand, listener modeling \cite{ldnet}, which was a widely used technique in VMC 2022, was rarely used this year. This is possibly due to the fact that the datasets of tracks 2 and 3 did not come with listener-wise labels.

As for the model architecture, most teams follow the design of SSL-MOS, i.e., attaching a simple linear or recurrent neural network layer to predict the final score on top of the feature extractor (in most cases SSL model). Some teams tried to use more complicated models, such as EfficientNetV2 \cite{efficientnetv2}. Finally, model ensembling was an effective approach, as they were used by top teams including T04 and T05.

\subsection{Description of top systems}

\noindent\textbf{T06.} The T06 system performed remarkably well in all three tracks. Specifically, they ranked first in 9 of the 16 metrics in track 1, ranked first in all four utterance-level metrics in track 2, and ranked first,  second, and first in SIG, BAK, and OVRL in track 3, respectively. Their system was an improved version of RAMP \cite{ramp}, whose core idea was to equip a parametric model (e.g., SSL-MOS) with a retrieval-based, non-parametric head based on kNNs. It was shown in \cite{ramp} that the retrieval-based method greatly improved the zero-shot generalization ability on unseen datasets. For this challenge, they further added a prior net to refine the distance distribution of the retrieved k neighbors. As for the dataset, they claimed that they only used the provided training sets.

\noindent\textbf{T05.} The T05 system ranked first in 7 of the 16 metrics in track 1. Their system used a wav2vec 2.0 feature extractor and an EfficientNetV2 encoder that took mel spectrogram as input, and a fusion network tried to predict the final output score. In addition to the provided datasets, they also used SOMOS, BC 2008, BC 2009, BC 2011. Moreover, they conducted their own listening test using the top half systems of BVCC, and used it as an additional dataset. They promised to publicize the dataset in the near future.

\noindent\textbf{T08.} The T08 system ranked first or second in five of the eight metrics in track 2. Their system was based on SSL-MOS, with an additional pitch histogram input, which reflects the fundamental frequency of each frame of a speech sample to the octave scale.

\noindent\textbf{T04.} The T04 system ranked third, first, and second in the prediction of SIG, BAK, and OVRL in track 3. They proposed an interesting approach: they trained two separate models for BAK and SIG prediction and simply took the average of the predicted BAK and SIG scores as the OVRL score. For the BAK prediction model, they pre-trained it to predict the SNR value of simulated noisy speech samples. The SNR values were mapped to a range of 1-5 to approximate the BAK scores (for instance, an SNR value of -20 dB was mapped to a BAK score of 1.0, an SNR of +50 dB was mapped to 4.5, and clean speech was mapped to 5.0). This pre-training step seems to be very useful as the LCC of BAK prediction is much higher than others. The SIG model was pre-trained using samples from the ASVSpoof 2019, where spoofed and natural samples were mapped to scores of 1 and 5, respectively. Finally, the two prediction models were fine-tuned on the provided training data.

\section{Feedback from the participants}
\label{sec:feedback}

We also collected feedback from the participants through the system description questionnaires. Overall, all participants found this challenge interesting, and the diverse task design this year reflects the need of this research field, especially for tasks 1 and 3. However, one participant was worried that having three tracks might be overwhelming for some potential participants. As for the data, most participants found the datasets of good quality. One participant found the audio samples in track 2 to be too short. Two participants found that the training and evaluation data of track 3 were different, where the former was from real-world environment, and the latter was simulated.

Regarding logistics, most participants found the CodaBench platform easy to use. One participant suggested providing a longer period between the release of the results and the paper submission deadline. Also, several participants suggested allowing multiple submissions.

Some participants also provided suggestions for future directions, including considering different sampling frequencies, synthetic speech from zero-shot voice cloning systems, SIG prediction, and other audio types such as music and environment sounds. One participant also suggested putting more attention on the utterance-level metrics. Finally, one participant suggested refining the BVCC dataset, considering that the samples (and consequently, the scores) are somewhat outdated.

\section{Conclusion}

We have presented a summary of the 2024 edition of the VoiceMOS Challenge. This year we provided three tasks, each reflecting an important need for the task of subjective speech evaluation prediction: prediction of high-quality speech synthesis systems, prediction of a diverse set of modern singing voice synthesis and conversion systems, as well as prediction of multiple aspects of noisy, clean and enhanced speech in a semi-supervised setting. Among the eight submissions we received, we confirmed that the baseline in each track was outperformed by at least one participating system, and we identified several emerging techniques that seemed to be effective.

As the generative AI technology for audio has been evolving at an all-time fast pace, the call for a timely evaluation method is desperate. The future directions suggested by the participants reflect this trend, as they include incorporating modern-day speech synthesis systems, more diverse speech types, or even extending the prediction target beyond speech to music or even environmental sounds.

\section{ACKNOWLEDGMENTS}
\label{sec:ack}

This work was partly supported by JST CREST under Grant Number JPMJCR19A3 and JST AIP Acceleration Research under Grant Number JPMJCR24U3. THis work was also parly supported by NSTC 112-2221-E-001-009-MY3.
\bibliographystyle{IEEEbib}
\bibliography{refs}

\end{document}